\documentclass[conference]{IEEEtran}
\usepackage{myStyleIEEE}
\usepackage{ushort}
\IEEEoverridecommandlockouts

\def\R{\mathbb{R}}
\def\C{\mathbb{C}}

\usepackage{siunitx}
\AtBeginDocument{\DeclareSIUnit{\MWh}{MWh}}
\AtBeginDocument{\DeclareSIUnit{\kWh}{kWh}}
\DeclareSIUnit \voltampere { VA } 
\DeclareSIUnit \var { var } 
\usepackage{amsmath}
\usepackage{mathtools}
\usepackage{isomath}
\let\underbrace\LaTeXunderbrace

\begin{document}

\title{Multiple Ancillary Services Provision by Distributed Energy Resources in Active Distribution Networks}

\renewcommand{\theenumi}{\alph{enumi}}

\newcommand{\uros}[1]{\textcolor{magenta}{$\xrightarrow[]{\text{U}}$ #1}}
\newcommand{\justin}[1]{\textcolor{blue}{$\xrightarrow[]{\text{J}}$ #1}}
\newcommand{\ognjen}[1]{\textcolor{red}{$\xrightarrow[]{\text{O}}$ #1}}

\author{
\IEEEauthorblockN{Ognjen Stanojev\IEEEauthorrefmark{1}, Yi Guo\IEEEauthorrefmark{1}, Petros Aristidou\IEEEauthorrefmark{2}, Gabriela Hug\IEEEauthorrefmark{1}}%
\IEEEauthorblockA{\IEEEauthorrefmark{1} EEH - Power Systems Laboratory, ETH Zurich, Switzerland } %
\IEEEauthorblockA{\IEEEauthorrefmark{2} Department of Electrical Engineering, Computer Engineering and Informatics, Cyprus University of Technology, Cyprus} %
Emails: \{stanojev, guo, hug\}@eeh.ee.ethz.ch, petros.aristidou@cut.ac.cy
\thanks{This research is supported by the Swiss National Science Foundation under NCCR Automation, grant agreement 51NF40\_180545.}
}

\maketitle
\IEEEpeerreviewmaketitle

\begin{abstract}
The electric power system is currently experiencing radical changes stemming from the increasing share of renewable energy resources and the consequent decommissioning of conventional power plants based on synchronous generators. Since the principal providers of ancillary services are being phased out, new flexibility and reserve providers are needed. The proliferation of Distributed Energy Resources (DERs) in modern distribution networks has opened new possibilities for distribution system operators, enabling them to fill the market gap by harnessing the DER flexibility. This paper introduces a novel centralized MPC-based controller that enables the concurrent provision of voltage support, primary and secondary frequency control by adjusting the setpoints of a heterogeneous group of DERs in active distribution grids. The input-multirate control framework is used to accommodate the distinct timescales and provision requirements of each ancillary service and to ensure that the available resources are properly allocated. Furthermore, an efficient way for incorporating network constraints in the formulation is proposed, where network decomposition is applied to a linear power flow formulation together with network reduction. In addition, different timescale dynamics of the employed DERs and their capability curves are included. The performance of the proposed controller is evaluated on several case studies via dynamic simulations of the IEEE 33-bus system.
\end{abstract}


\section{Introduction} \label{sec:intro}

Power system ancillary services are essential for the secure and reliable operation of transmission and distribution networks and have traditionally been provided by conventional power plants based on Synchronous Generators (SGs). However, a significant portion of the conventional generation units are expected to be decommissioned and replaced by Renewable Energy Sources (RES) in an effort to decarbonize the energy system. The large-scale integration of converter-interfaced RES imposes new challenges on real-time control and operation, as the lack of rotational inertia and damping leads to faster dynamics and larger frequency deviations, which can adversely affect the overall system stability \cite{Milano2018}. The consequences expand even further, with over-voltages and thermal overloads being more likely to occur, especially in high RES production and low load conditions. The above-mentioned developments elucidate the need for new flexibility and ancillary service providers.

Modern active Distribution Networks (DNs) are populated with a vast number of Distributed Energy Resources (DERs), such as Battery Energy Storage Systems (BESS), Diesel Generators (DG), Photovoltaics (PVs), and Flexible Loads (FLs), and thus contain a substantial amount of operational flexibility, which is still an untapped resource. Therefore, multitudes of DERs can be aggregated and collectively controlled to provide ancillary services through regulation of the DN's power exchange with the transmission system\cite{Hatziargyriou2017}. Aggregation strategies, including virtual power plants \cite{DallaneseVPPRegulation2018}, DER clusters \cite{Tang2018}, and load aggregators \cite{TangLoad2018}, have recently been proposed to effectively harness the collective flexibility from a large number of DERs. To this end, sufficient capacity of flexible units is aggregated for active participation of DNs in energy and ancillary service markets, and at the same time, the security of supply in the local network is ensured. Paradigms described above rely on the availability of communication infrastructure for network management, which is typically present in modern, moderate scale DNs \cite{MGControl2014}. Furthermore, efficient real-time algorithmic frameworks need to be developed for such strategies to become a reality.

The provision of ancillary services by DERs in distribution grids has recently been the subject of many publications. Thus far, services such as voltage support \cite{Valverde2019,Aristidou2017}, primary \cite{Vrettos2013,Koller2015,OgnjenPowerTech2021} and secondary \cite{Galus2011,Tang2018} frequency control, congestion management \cite{Knezovic2015,Contreras2021}, etc., have been studied. In \cite{Valverde2019}, the provision of voltage support to the transmission system from small-scale PV systems hosted in distribution networks by means of their local control has been demonstrated.  The work in \cite{Tang2018} considered a distributed algorithm for tracking the frequency control signal by a cluster of DERs, whereas \cite{Contreras2021} developed an aggregation method to employ flexibility from DERs for management of congestion in the transmission grid. Although earlier works in this context considered only single DER types, e.g. BESS \cite{Koller2015}, thermostatically controlled loads \cite{Vrettos2013} or small-scale PVs \cite{Valverde2019}, more recent literature \cite{StavrosAS2020,DallaneseVPPRegulation2018} acknowledges the importance of aggregation of multiple DER types by considering specific limitations and requirements for each unit type. Moreover, the majority of works \cite{Galus2011,Vrettos2013,Koller2015} disregard the line flow and nodal voltage constraints, and thus, a single connection point to the transmission system for all DERs is assumed. However, the range of feasible operating points is limited by static and dynamic properties of the network components and operational circumstances, which need to be taken into account in the control design.

Ancillary services can be provided by forcing the power flow at the point of common coupling to follow a reference signal commanded by the Transmission System Operator (TSO), as has been studied in \cite{DallaneseVPPRegulation2018,Karthikeyan2019,Mayorga2017}. In \cite{DallaneseVPPRegulation2018}, a distributed optimization framework leveraging online primal-dual-type methods was developed to control the output powers of DERs. In \cite{Karthikeyan2019}, a Model Predictive Control (MPC)-based controller is developed for the utilization of flexible resources such as BESS and FLs to provide demand response by adjusting the power flow at the feeder head in low-voltage grids. Although efficient algorithms for the main feeder power flow regulation are proposed by the aforementioned references, specific ancillary services were not considered and thus, the problem of ancillary services provision is simplified. In addition, dynamic properties of the system frequency and DERs are typically neglected, as most works resort to steady-state modeling approaches. The significance of including DN dynamics was emphasized in \cite{Mayorga2017}, where a rule-based controller for the provision of ancillary services was developed. 

In all of the aforementioned studies, only one specific ancillary service is considered. While \cite{StavrosAS2020} considers provision of primary, secondary and tertiary frequency control, only a single ancillary service is offered at each time. Similar to conventional power plants, active DNs will be able to participate in multiple ancillary services simultaneously in the future, and therefore, the development of control schemes that will enable integration of multiple ancillary services within a single formulation is becoming increasingly important. Conventionally, control structures for regulating voltage levels or frequency are realized via a number of nested control loops, decoupled from each other and activated in a cascaded fashion \cite{ULBIG2011}. Due to time-scale separation and typically sufficient transmission line capacities, interdependencies and interactions between different control layers are not considered in the design of individual-level controls. However, distribution grids operate under significantly smaller line capacity and allowable voltage deviation margins. Therefore, concurrent provision of multiple ancillary services by DERs requires integration of various control levels into a single formulation, where sharing of the DER resources as well as network capacities between the individual control levels is considered. 

This paper proposes a centralized multirate MPC-based controller that adjusts the power setpoints of various DERs in active DNs in response to frequency and voltage deviations at the Point of Common Coupling (PCC) to provide voltage support, as well as primary and secondary frequency control. In contrast to \cite{StavrosAS2020}, where different ancillary services were considered independently, we establish a unified framework capable of accommodating multiple ancillary services at the same time. To this end, an input-multirate control framework \cite{Ravi2009} is employed to include distinct time scales and provision requirements of each individual ancillary service as well as to ensure that the available resources are properly allocated. Using this framework, the MPC problem is formulated as a constrained linear periodic system with time-varying dimensions and a quadratic objective function. An efficient way of incorporating network constraints in the formulation is proposed, where network decomposition is applied to a linear power flow formulation together with network reduction. Thus, the model size is reduced with only a minor loss of accuracy. Furthermore, different timescale dynamics of the employed DERs and their capability curves are included in the formulation. Finally, contrary to other studies, e.g. \cite{Tang2018, StavrosAS2020}, the proposed control design is verified through time-domain simulations of the IEEE 33-bus system with detailed dynamic models of loads, network lines, and DER units.

The rest of the paper is organized as follows. Section~\ref{sec:ASprovision} summarizes the basic properties of ancillary services considered in this work and presents the proposed control scheme. The model predictive control algorithm used to dispatch the DERs is discussed in Sec.~\ref{sec:OPF}, with its final form presented in Sec.~\ref{subsec:mpcfinal}. The input-multirate control framework, which enables integration of multiple ancillary services into a single formulation is introduced in Sec.~\ref{subsec:imcf}, and a linear network model suitable for real-time control is proposed in Sec.~\ref{subsec:BFSdecomposed}. Finally, the performance of the proposed controller is tested by performing multiple case studies in Sec.~\ref{sec:res}.  

\section{Active DNs Providing Ancillary Services} \label{sec:ASprovision}
In this section, we first briefly review the rules and principles of ancillary services considered in this study and subsequently give an overview of the proposed centralized control scheme. Regulations and properties for the provision of control reserves vary substantially between countries and control areas. Therefore, we adopt standard guidelines and conventions for the Continental Europe system (ENTSO-E) established in \cite{entsoe_FSM}. Within this framework, active DNs with controlled DER aggregations fall into the category of Type C power-generating modules, which provide principal ancillary services to ensure the security of supply.

\subsection{Primary Frequency Control}
The Primary Frequency Control (PFC) services are activated by a decentralized proportional controller within the governors of the responsible units. The needed PFC power $P_\mathrm{pfc}\in\R$ depends on the frequency deviation $\Delta f\in\R$ from the nominal $\SI{50}{\Hz}$ and the contracted amount of up- and down-regulation $(\overline{P}_\mathrm{pfc},\underline{P}_\mathrm{pfc})\in\R^2$, and can be expressed via the following droop expression:
\begin{equation}
    \label{eq:pfc}
    P_\mathrm{pfc}(\Delta f) = \begin{cases}
    \min(\overline{P}_\mathrm{pfc}, k_{p,p} \cdot \Delta f), & \Delta f \geq 0\\
    \max(\underline{P}_\mathrm{pfc},k_{p,p} \cdot \Delta f), & \Delta f < 0
    \end{cases},
\end{equation}
where $k_{p,p}\in\R_{\geq0}$ represents the droop gain. In Continental Europe, primary frequency control is designed to be a symmetric product, i.e., $\overline{P}_\mathrm{pfc}=\underline{P}_\mathrm{pfc}$. It has to be fully activated within $\SI{15}{\second}$ and sustained for the maximum duration of $\SI{15}{\min}$. Although ENTSO-E recommends a narrow deadband to be implemented, it is not considered in this work for simplicity.

\subsection{Secondary Frequency Control}
The Secondary Frequency Control (SFC) reserves are initiated by a proportional-integral controller operated by the TSO to relieve the primary control reserves and restore the system frequency to its nominal value while ensuring that the scheduled tie-line exchanges with other control areas are maintained. More precisely, the area control error $e_a\in\R$ is minimized by means of PI control $(k_{p,a},k_{i,a})\in\R^2_{\geq0}$:
\begin{equation}
     P_\mathrm{sfc}(s) = (k_{p,a}+\frac{k_{i,a}}{s})\cdot \underbrace{(\Delta p_{t} + B\Delta f)}_{e_a}, \label{eq:sfc}
\end{equation}
where $\Delta p_{t}\in\R$ represents the deviation from the scheduled tie-line exchange with other control areas, and $B\in\R_{>0}$ is the bias factor of the control area. The secondary control signal is then formed by weighting the required power adjustment $P_\mathrm{sfc}(s)\in\C$ and transmitting it to the providing units. Typical response times for the secondary control activation and deployment are in the range from \SI{30}{\second} to \SI{15}{\min}.

\subsection{Transmission Network Voltage Support}
Voltages in the transmission system are maintained within safe limits by means of Voltage Control (VC), which is activated by a proportional controller within voltage regulators of generating units when a variation in voltage $\Delta v\in\R$ across the providing unit's terminal is detected. Similarly to PFC, reactive power injection $Q_\mathrm{vc}\in\R$ needed for voltage control is calculated by the following droop equation:
\begin{equation}
    \label{eq:vc_q}
    Q_\mathrm{vc}(\Delta v) = \begin{cases}
    \min(\overline{Q}_\mathrm{vc}, k_{p,v} \cdot \Delta v), & \Delta v \geq 0\\
    \max(\underline{Q}_\mathrm{vc},k_{p,v} \cdot \Delta v), & \Delta v < 0
    \end{cases},
\end{equation}
where the contracted up- and down-regulation is denoted by $(\overline{Q}_\mathrm{pfc},\underline{Q}_\mathrm{pfc})\in\R^2$ and $k_{p,v}\in\R$ is the droop gain. The response time is usually between several milliseconds and one minute. Active participation in voltage support has recently become mandatory for distribution networks \cite{StavrosVC}.

\subsection{Proposed Control Structure}
This work considers a radial balanced DN represented by a connected graph $\mathcal{G}=(\mathcal{N},\mathcal{E})$, with $\mathcal{N} \coloneqq \{0,1,\dots,N\}$ denoting the set of network nodes including the substation node $0$, and $\mathcal{E} \subseteq \mathcal{N}\times\mathcal{N}$ designating the set of $N$ network branches. The distribution network hosts a number of DERs and loads, where $\mathcal{D}\subseteq \mathcal{N}$ indicates the subset of nodes with DGs, $\mathcal{P}\subseteq \mathcal{N}$ the subset of nodes with PVs, $\mathcal{B}\subseteq \mathcal{N}$ the subset of nodes with BESSs, $\mathcal{L}\subseteq \mathcal{N}$ the subset of nodes with loads, and $\mathcal{C}\subseteq\mathcal{L}\subseteq \mathcal{N}$ is the subset of nodes with FLs. Variable Speed Heat Pumps (VSHPs) are considered as a representative of the FLs. The set of nodes with DERs is thus obtained by the following union of sets: $\mathcal{R} \coloneqq \mathcal{D} \cup \mathcal{P} \cup \mathcal{B} \cup \mathcal{C}$. Cardinality of the previously defined sets is denoted by: $n_d \coloneqq |\mathcal{D}|$, $n_p \coloneqq |\mathcal{P}|$, $n_b \coloneqq |\mathcal{B}|$, $n_l \coloneqq |\mathcal{L}|$, $n_c \coloneqq |\mathcal{C}|$, $n_g \coloneqq |\mathcal{R}|$.

\begin{figure}[!t]
    \centering
    \includegraphics[scale=0.9]{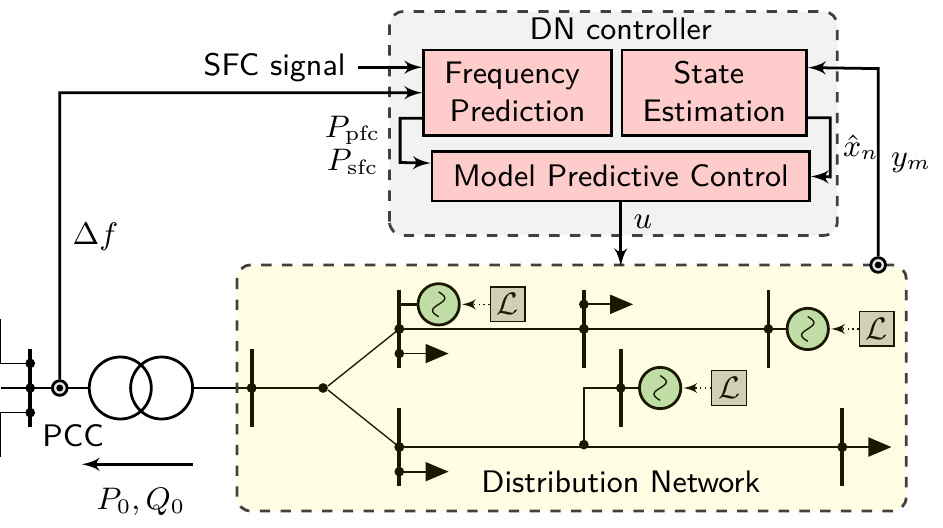}
    \caption{Proposed active DN control structure.}
    \label{fig:control_structure}
    \vspace{-0.35cm}
\end{figure}

The proposed control structure is depicted in Fig.~\ref{fig:control_structure}, where an active DN is centrally controlled to provide ancillary services by regulating the power exchange $(P_0,Q_0)\in\R^2$ with the transmission grid. In a normal operating state, the DERs are regulated by their respective local controllers $\mathcal{L}$, with the principal objective of ensuring grid synchronization and reference tracking. However, in events of frequency and voltage deviations, the local controller setpoints are adjusted by the central DN controller such that the active and reactive power required for the transmission network support is delivered. The central controller uses the available measurements, both from the local grid and the PCC, to compute the optimal setpoint adjustments and is composed of a state estimator, a frequency prediction module and an MPC algorithm. 

\subsubsection{State Estimation} Measurements available across the DN such as bus voltages, branch currents and DER outputs are described by vector $y_m \in\R^{N_y}$, with $N_y\in\N$ being the number of measurements. The measurement noise $\xi\in\R^{N_y}$ is assumed to follow a normal distribution with zero mean and covariance $\Sigma\in\R^{N_y\times N_y}$. Thus, the measurement model can be represented by $y_m = h(x_n) + \xi$, with $h:\R^{2N}\xrightarrow{}\R^{N_y}$ denoting a linear measurement mapping. A Weighted Least Squares state estimation is employed to process the obtained measurements and determine the state of the DN $\hat{x}_n\in\R^{2N}$, as follows:
\begin{equation} 
    \hat{x}_n = \underset{x_n}{\textrm{argmin}}\,\,\, \frac{1}{2}\left(y-h(x_n)\right)^\mathsf{T}W_\mathrm{SE}\left(y-h(x_n)\right), \label{eq:WLS_SE}
\end{equation}
where $W_\mathrm{SE}\coloneqq(\Sigma)^{-1}$ is the weight matrix. 
Since the focus of this work is not on DN estimation techniques, we assume that sufficient measurements are available to guarantee full observability of the DN. For more detailed analysis and discussion on state estimation of distribution grids, we refer the reader to \cite{Primadianto2017}.

\subsubsection{Frequency Prediction Model}
Given that MPC is the controller of choice in the proposed approach, a prediction of the active power to be delivered for the PFC provision $P_\mathrm{pfc}\in\R$ is needed, which on the other hand requires a prediction of the system frequency for the future time period. We employ a model to predict the center-of-inertia frequency of a low-inertia system proposed in \cite{UrosLQR}, where the relationship between the system frequency deviation $\Delta f(s)\in\C$ and a change in the power balance $\Delta p(s)\in\C$ is represented by a second-order transfer function:
\begin{equation}
    G(s) = \frac{\Delta f(s)}{\Delta p_e(s)} = \frac{1}{MT}\frac{1+sT}{s^2+2\zeta\omega_n s + \omega_n^2}, \label{eq:G}
\end{equation}
with the natural frequency $\omega_n\in\R_{>0}$ and damping ratio $\zeta\in\R_{>0}$ computed as follows: 
\begin{equation}
    \omega_n = \sqrt{\frac{D+R_g}{MT}}, \quad \zeta = \frac{M+T(D+F_g)}{2\sqrt{MT(D+R_g)}}. \label{eq:wn}
\end{equation}
Here, the parameters $R_g\in\R_{>0}$ and $F_g\in\R_{>0}$ denote the average inverse droop control gain and the fraction of total power generated by the high-pressure turbines of SGs, $T\in\R_{>0}$ represents the generator time constant, and $M\in\R_{>0}$ and $D\in\R_{>0}$ designate the weighted system averages of inertia and damping constants, respectively. Hence, the model considers both synchronous and inverter-based generation, with the inclusion of inertial response and primary frequency control. Given a stepwise disturbance in the electrical power $\Delta p_e(s) = -\Delta P/s$, the state-space representation of the following form is obtained: 
\begin{equation}
\label{eq:ss_general}
\begin{bmatrix}
    \dot{\omega} \\ 
    \ddot{\omega} 
\end{bmatrix}
= 
\begin{bmatrix}
    0 & 1 \\
    -\frac{D+R_g}{MT} & -(\frac{1}{T}+\frac{D+F_g}{M})
\end{bmatrix} 
\begin{bmatrix}
    \omega \\ 
    \dot{\omega}
\end{bmatrix}
+
\begin{bmatrix}
    0 \\ \frac{\Delta P}{T M}
\end{bmatrix},
\end{equation}
where $\Delta P\in\R$ represents the power imbalance magnitude, $x_f\coloneqq \begin{bmatrix} \omega & \dot{\omega}\end{bmatrix}^\mathsf{T}\in\R^2$ denotes the state vector, and $\omega \equiv \Delta f(t)\in\R$ designates the frequency deviation. The state vector initial value $x_f(0)$ can be retrieved at each time step from the frequency and RoCoF measurements at the PCC. Furthermore, the power imbalance magnitude $\Delta P$ can be determined either locally from RoCoF measurements, as discussed in \cite{OgnjenMPC2020}, or using data-driven algorithms described in \cite{OgnjenSEST2020}. The discrete-time form of the state-space model \eqref{eq:ss_general} is obtained by applying the zero-order hold method and used for the frequency evolution prediction. Note that in the setup presented in Fig.~\ref{fig:control_structure} the frequency prediction module is external to the MPC procedure, which is justified when the DN cannot significantly impact the frequency dynamics due to its limited capacity. Otherwise, the prediction module should be included within the MPC procedure, as has been done in \cite{OgnjenMPC2020}.

Lastly, outputs of the frequency prediction module and the state estimator are passed to the MPC algorithm to compute the optimal DER setpoint adjustments $u\in\R^{2 n_g}$. Active and reactive power exchanges with the transmission grid $(P_0,Q_0)$ are thus forced to deviate from the scheduled values such that the considered ancillary services are provided. 

\section{Multirate Model Predictive Control} \label{sec:OPF}
This section is devoted to the formulation of the centralized MPC algorithm, which is designed to accommodate the provision of the three ancillary services introduced in the previous section, i.e., primary and secondary frequency control, and voltage control. To this end, two control input update rates are considered: (i) a fast rate related to the execution of primary frequency and voltage control actions, and (ii) a slower rate related to tracking of the SFC control signal from the TSO. Additional ancillary services, e.g. tertiary frequency control, can be included following the same methodology by adding another update rate.

\begin{figure}[!b]
    \centering
    \includegraphics[scale=0.975]{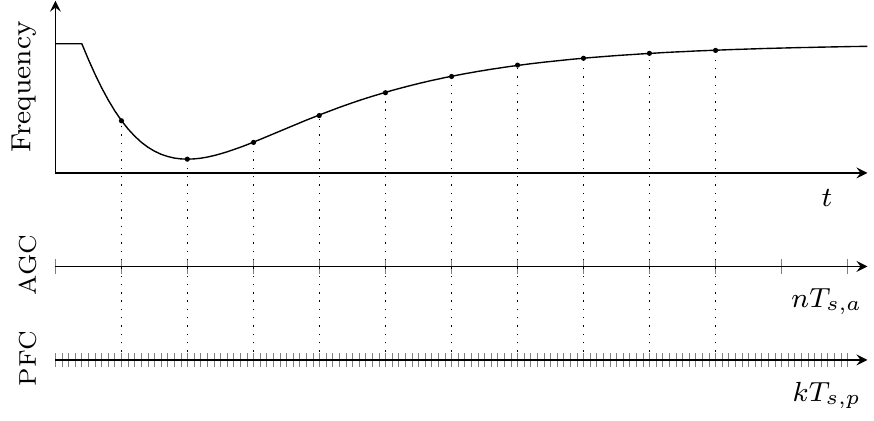}
    \caption{An exemplary timing diagram indicating differences in update rates between AGC and PFC (VC) inputs. Time instances when the entire control input vector is updated are marked with a dashed vertical line.}
    \label{fig:timing_diag}
\end{figure}
\subsection{Input-Multirate Control Framework} \label{subsec:imcf}
Control actions associated with the provision of primary frequency and voltage control in the form of DER active and reactive power setpoint changes $(\Delta P^{\star}_{p,k},\Delta Q^\star_{p,k})\in\R^{2n_g}$ are executed at time instants $t_k\in\{kT_{s,p}\}_{k\in\mathbb{N}}$, where $T_{s,p}\in\R_{>0}$ is the time required to compute and broadcast the control inputs to the individual DERs. On the other hand, control actions pertaining to provision of the secondary frequency control $\Delta P^{\star}_{a,k}\in\R^{n_g}$ are applied at time instants $t_n\in\{nT_{s,a}\}_{n\in\mathbb{N}}$ with a larger time step $T_{s,a}\gg T_{s,p}$, due to a slower rate at which the SFC providers are required to react and the longer times needed to transmit the SFC signal. The ratio between the two sampling periods is assumed to be an integer value and is denoted by $q\coloneqq T_{s,p}/T_{s,a}\in\mathbb{N}$. This setup corresponds to the input-multirate control framework\cite{Ravi2009}, where each input channel has a unique sampling period. Understanding of the concept can be aided with the diagram presented in Fig.~\ref{fig:timing_diag}. Within this framework, all input channels are initialized synchronously, i.e., $t_{k=0}$ corresponds to $t_{n=0}$. Input $\Delta P_{a,k}^\star$ can be updated only at time steps $k\in\mathcal{U}\coloneqq\{k\in\N:\mathrm{mod}(k,q)=0\}$, while at all other time steps the input remains unchanged, i.e., $\Delta P_{a,k}^\star=\Delta P_{a,k-1}^\star$. Let us furthermore define $z$ to be a binary variable indicating if all channels are updated at time step $k$, and $m_z\in\mathbb{N}$ to denote the number of decision variables updated at time step $k$:
\begin{equation}
    z=\begin{cases}
    1, & \mathrm{if}\,\, \mathrm{mod}(k,q) = 0\\
    0, & \mathrm{if}\,\,\mathrm{mod}(k,q) \neq0
    \end{cases},\,
    m_z=\begin{cases}
    3n_g, &\mathrm{if}\,\, z = 1\\
    2n_g, &\mathrm{if}\,\, z = 0
    \end{cases}.
\end{equation}
The controller sampling period corresponds to that of the faster input channel ($T_s\coloneqq T_{s,p}$), but the entire control input vector $u_k\coloneqq(\Delta P^{\star}_{k},\Delta Q^\star_{k})\in\R^{2n_g}$ is updated solely at time steps which are integer multiples of $q$, and can be represented as:
\begin{equation} 
        u_k = C_z \bar{u}_k + \hat{u}_k, \label{eq:u_def}
\end{equation}
where $\bar{u}_k\in\R^{m_z}$ contains all input channels which are updated at step $k$, and $\hat{u}_k\in\R^{3n_g-m_z}$ collects all input channels which remain constant at time step $k$. More precisely, the variables and matrices in \eqref{eq:u_def} are defined for every time step $k$ as follows:
\scalebox{0.92}{\parbox{1.08\linewidth}{%
\begin{subequations} \label{eq:Cz}
\begin{align}
    \bar{u}_k & = \begin{bmatrix}
        \Delta P^\star_{p,k} \\ \Delta Q^\star_{p,k} \\            \Delta P^\star_{a,k}
    \end{bmatrix},
    C_z = \begin{bmatrix}
        \mathbbl{1}_{n_g} & \mathbbl{0}_{n_g} \\ 
        \mathbbl{0}_{n_g} & \mathbbl{1}_{n_g} \\
        \mathbbl{1}_{n_g} & \mathbbl{0}_{n_g}
    \end{bmatrix}^\mathsf{T}, \hat{u}= 0,\,\, \forall k\in\mathcal{U}, \\
    \bar{u}_k & = \begin{bmatrix}
       \Delta P^\star_{p,k} \\ \Delta Q^\star_{p,k}
    \end{bmatrix},\,\, C_z = \mathbbl{1}_{2n_g},\,\, \hat{u}_k = \Delta P_{a,k}^\star,\,\, \forall k\in\mathbb{N}/\mathcal{U},
\end{align}
\end{subequations}
}}
with $\mathbbl{1}_{n}$ representing an identity matrix of size $n\in\mathbb{N}$, and $\mathbbl{0}_{m}$ denoting a zero matrix of size $m\in\mathbb{N}$. 
Note that the control input vector $u_k$ has constant dimensions while dimensions of vectors on the right-hand side of \eqref{eq:u_def} are time-varying.

\subsection{Objective Function}
The control goal of minimizing the total control effort associated with the provision of ancillary services as well as the network losses, over all time steps $\mathcal{H}\coloneqq\{\hat{k}, \hat{k}+1, \dots,\hat{k}+H\}$ within the prediction horizon of length $H$, is reflected in the following objective function:
\begin{equation} \label{eq:objective_fcn}
\begin{split}
\underset{\bar{u}_{k,k\in\mathcal{H}}}{\min}\,\,\,&\sum_{k\in\mathcal{H}}\sum_{d\in\mathcal{R}}\Big(C_{P_d}(\Delta P_{p,d,k}^\star)^2 + C_{Q_d}(\Delta Q_{p,d,k}^\star)^2 +\\& C_{A_d}(\Delta P_{a,d,k}^\star)^2 \Big) + \sum_{k\in\mathcal{H}}I_k^\mathsf{T}R  I_k,
\end{split}
\end{equation}
where $\Delta P^\star_{p,d,k}\in\R$ and $\Delta Q^\star_{p,d,k}\in\R$ are the active and reactive power setpoint changes related to the provision of PFC and VC of the DER connected at node $d$ at time step $k$, $\Delta P^\star_{a,d,k}\in\R$ is similarly the active power setpoint change related to provision of SFC of the unit connected at node $d$ at time step $k$, $R\in\R^{N\times N}_{\geq0}$ is a diagonal matrix populated with branch resistances, and $I_k\in\C^N$ is a vector collecting the branch currents at time step $k$. The cost coefficients $C_{P_d}\in\R_{\geq 0}$, $C_{Q_d}\in\R_{\geq 0}$ and $C_{A_d}\in\R_{\geq 0}$ are selected based on the following relationship:
\begin{equation}
    C_{Q_d} \leq C_{P_d} \leq C_{A_d},\,\,\,\forall d\in\mathcal{R},
\end{equation}
which prioritizes ancillary services related to stability, i.e. voltage and primary frequency over secondary frequency control, denoted respectively by subscripts $Q$, $P$ and $A$. Furthermore, dispatch priorities of different DER types are also enforced by considering battery deterioration, fuel costs, and PV and flexible load curtailment prices:
\scalebox{0.88}{\parbox{1.13\linewidth}{%
\begin{equation}
    C_{X_p} \leq C_{X_b} \leq C_{X_v} \leq C_{X_g}, \forall p\in\mathcal{P}, \forall b\in\mathcal{B}, \forall g\in\mathcal{D}, \forall v\in\mathcal{V},
\end{equation}
}}
where $X\in\{P,Q,A\}$ is used to denote the appropriate ancillary service.

\subsection{Dynamics of Distributed Energy Resources}
To improve accuracy and optimality of the control design, relevant dynamics of DERs are taken into account as a part of the DN model. Thus, in this work, we model ramping dynamics of DGs and VSHPs as well as the state-of-charge dynamics of batteries. The dynamics pertaining to power ramping of BESS and PV units are assumed to have significantly smaller time constants compared to the controller sampling period. Therefore, any change in the setpoint values $(\Delta P^\star_{d,k},\Delta Q^\star_{d,k})\in\R^2$ is assumed to be directly reflected in the active and reactive power outputs:
\begin{align} \label{eq:bess_p}  
    P_{d,k+1} &= P_{d,k}^\star + \Delta P_{d,k}^\star, \quad\, \forall d\in\mathcal{P}\cup\mathcal{B}, \\ 
    Q_{d,k+1} &= Q_{d,k}^\star+\Delta Q_{d,k}^\star, \quad \forall d\in\mathcal{P}\cup\mathcal{B},
\end{align}
where $P_{d,k}\in\R$ and $Q_{d,k}\in\R$ denote the active and reactive power outputs of the appropriate unit $d\in\mathcal{P}\cup\mathcal{B}$ at time step $k$, and similarly, $P_{d,k}^\star\in\R$ and $Q_{d,k}^\star\in\R$ are the active and reactive power setpoints. 
\subsubsection{DG Dynamics}
The governor and the exciter dynamics of DG units are modeled by a discrete first order filter as:
\scalebox{0.89}{\parbox{1.1\linewidth}{%
\begin{align}
    P_{g,k+1} &=a_{P_g}(P_{g,k}^\star + \Delta P_{g,k}^\star) + (1-a_{P_g})P_{g,k},\,\,\,\, \forall g\in\mathcal{D}, \\
    Q_{g,k+1} &=a_{Q_g}(Q_{g,k}^\star+\Delta Q_{g,k}^\star) + (1-a_{Q_g})Q_{g,k}, \forall g\in\mathcal{D},
\end{align}
}}

\noindent
with $P_{g,k}\in\R$ and $Q_{g,k}\in\R$ representing the DG active and reactive power outputs at time step $k$, $P_{g,k}^\star\in\R$ and $Q_{g,k}^\star\in\R$ denoting the active and reactive power setpoints, $a_{P_g}\coloneqq 1-e^{-T_s/T_g}\in\R_{\geq 0}$ being a constant defined by the controller sampling period $T_s\in\R_{\geq 0}$ and the governor time constant $T_g\in\R_{\geq 0}$ specific to each unit $g$. Correspondingly, $a_{Q_g}\coloneqq 1-e^{-T_s/T_{e_g}}\in\R_{\geq 0}$ is a constant defined by the controller sampling period and the exciter time constant $T_{e_g}\in\R_{\geq 0}$.

\subsubsection{BESS State-of-Charge}
The state-of-charge $\chi_{b,k}\in\R$ of every BESS $b\in\mathcal{B}$ at each time step $k$ is modelled as
\begin{equation}
    \chi_{b,k+1} = \chi_{b,k} - T_s \cdot \frac{P_{b,k}^\star+\Delta P_{b,k}^\star}{E_{b}}, \,\,\,\, 
\end{equation}
with $P_{b,k}^\star\in\R$ denoting the BESS active power setpoint at time step $k$, and $E_b \in\R_{\geq 0}$ being the battery energy capacity. For simplicity, the battery storage is assumed to be lossless.
\subsubsection{VSHP Dynamics} A third-order state-space formulation is used to represent the VSHP active power $P_{v,k}\in\R$ consumption at time step $k$ in response to a setpoint change $\Delta P^\star_{v,k}$, as follows:
\begin{align}
    \nu_{v,k+1} &= 
    \begin{bmatrix}
        0 &T_s &0 \\ 
        0 &0 &T_s \\
        a_3 &a_2 &a_0
    \end{bmatrix} \nu_{v,k} + 
    \begin{bmatrix}
        T_s \\ 0 \\ 0
    \end{bmatrix}
 \Delta P_{v,k}^\star, \forall v\in\mathcal{V}\\
    P_{v,k+1} &= \begin{bmatrix}
        b_2 &b_1 &b_0
    \end{bmatrix} \nu_{v,k}, \qquad\qquad\qquad\quad\,\,\forall v\in\mathcal{V} \label{eq:vshp_p}
\end{align}
where $\nu_{v,k}\in\R^3$ is the state vector, $(a_1,a_2,a_3)\in\R^3$ and $(b_0,b_1,b_2)\in\R^3$ are constant coefficients that define the active power dynamics and can be obtained through transfer function fitting \cite{Johanna2020}. It is assumed that the VSHP is operating with constant power factor $\phi_v\in[0,\pi]$ and thus, the reactive power consumption is defined by $Q_{v,k}=P_{v,k}\tan(\arccos(\phi_v))$.

\subsection{Capability Curves of Distributed Energy Resources}
The active and reactive power outputs of DERs need to comply with their hardware and operational limitations represented by the so-called capability curves. Typically, the capability curves are defined as sets of allowable setpoints, as follows:
\begin{equation}
    (P_{d,k}^\star+\Delta P_{d,k}^\star,Q_{d,k}^\star + \Delta Q_{d,k}^\star) \in\mathcal{F}_{d}, \forall k\in\mathcal{H}, \forall d\in\mathcal{R}, \label{eq:capability_curves}
\end{equation}
where $P_{d,k}^\star\in\R$ and $Q_{d,k}^\star\in\R$ are active and reactive power setpoints of the respective unit $d$ at time step $k$, and the capability curves $\mathcal{F}_d = \{x \in \mathbb{R}^2 :\, A_d x \leq b_d\}$ are modeled as polytopes with $m_d\in\mathbb{N}$ edges defined by $A_d \in \mathbb{R}^{m_d \times 2}$ and $b_d \in \mathbb{R}^{m_d\times 1}$. The operating region of the DGs is limited by the stator current limit and a minimum active power output, the PV operating region is defined by the minimum power factor constraint and apparent power limit, the BESS capability curve is defined by its possibility of four-quadrant operation, and finally, the VSHP operation is limited by the constant power factor and minimum and maximum power output values. 

\subsection{Linear Power Flow Constraints Based on the BFS Method} \label{subsec:BFSdecomposed}
In this work, we employ the Backward/Forward Sweep (BFS) method proposed in \cite{Teng2003} for modeling the power flow, due to its computational efficiency, user-friendliness, and extensibility to unbalanced \cite{StavrosUnbalanced}, weakly meshed, and multi-phase grids \cite{Teng2003}. The method consists of a linearization step, where the nodal current injections $I^\mathrm{inj}_k \in\C^{N}$ are computed based on active $P^\mathrm{inj}_k\in\R^{N}$ and reactive $Q^\mathrm{inj}_k\in\R^{N}$ power injections; the backward sweep, where the branch currents $I_k\in\C^N$ are calculated using these current injections $I^\mathrm{inj}_k$; and the forward sweep, where the voltage drops over all branches $\Delta V_k\in\C^N$ are determined. Finally, nodal voltages $V_k\in\C^N$ are updated based on the computed voltage drops and the process is repeated until convergence, with the newest voltage updates used in the linearization step in every iteration. A single BFS iteration represents a linearized power flow model and can be described at every time step $k$ as follows:
\begin{align}
    I_k^\mathrm{inj} &= \mathrm{diag}(1/\bar{V}^*)
            \begin{bmatrix}
                 \mathbbl{1}_{N} & \mathbb{J}_{N}^* 
            \end{bmatrix}
            \begin{bmatrix}
                P^\mathrm{inj}_k \\
                Q^\mathrm{inj}_k
            \end{bmatrix}, \\
    I_k &= BIBC \cdot I^\mathrm{inj}_k, \label{eq:bfs_branch_model}\\
    \Delta V_k &= BCBV \cdot I^\mathrm{br}_k, \\
    V_k &= V_{s} - \Delta V_k, \label{eq:bfs_voltage_model} 
\end{align}
where $\bar{V}^*$ is the complex conjugate of \textit{a priori} determined nodal voltages, $V_{s}\in\R^N$ is a column vector of per-unit slack bus voltage magnitudes, and $\mathbb{J}_N$ is a diagonal matrix of size $N$ with imaginary unit $j\coloneqq\sqrt{-1}$ populating the diagonal entries. Furthermore, $BIBC \in \{0,1\}^{N\times N}$ represents a matrix of ones and zeros capturing the network topology, and $BCBV \in\C^{N\times N}$ is a complex matrix of the network branch impedances. The matrix $DLF\coloneqq BCBV\cdot BIBC$, which establishes the relationship between nodal current injections and voltage drops, is also commonly used. Using only a single iteration of the BFS algorithm for modeling power flows was previously shown to be a valid approach for including a linear network representation in an optimal power flow setup \cite{StavrosAS2020}, thus avoiding nonlinearities otherwise introduced by the AC power flow equations.

Taking advantage of the linearity of the formulation, the model is further simplified by using the superposition principle. It states that the response in any branch or node of a linear circuit having more than one independent source equals the sum of the responses caused by each independent source alone, with all other independent sources replaced by their internal impedances \cite{Hughes2008}. A similar procedure was used in \cite{Azizi2021} for localization of generator loss in the transmission system. Namely, consider a change in DER active and reactive power outputs that drive the network from its initial to a new state. Let $I^\mathrm{inj}_\mathrm{post}=I^\mathrm{inj}_\mathrm{pre} + \Delta I^\mathrm{inj}$ represent the vector of new current injections, where $I^\mathrm{inj}_\mathrm{pre}\in\R^N$ are the initial current injections and $\Delta I^\mathrm{inj}\in\R^N$ is the vector populated with current injection adjustments resulting from the DER output changes. By substituting $I^\mathrm{inj}_\mathrm{post}$ into \eqref{eq:bfs_branch_model}-\eqref{eq:bfs_voltage_model} we obtain: 

{\scalebox{0.88}{
\begin{minipage}{.55\linewidth}
\begin{subequations} \label{eq:base_bfs}
\begin{align}
    I_\mathrm{pre} &= BIBC \cdot I^\mathrm{inj}_\mathrm{pre}, \\
    V_\mathrm{pre} &= V_s - DLF \cdot I^\mathrm{inj}_\mathrm{pre},
\end{align}
\end{subequations}
\end{minipage}%
\begin{minipage}{.52\linewidth}
\begin{subequations} \label{eq:delta_bfs}
\begin{align}
    \Delta I &= BIBC \cdot \Delta I^\mathrm{inj}, \\
    \Delta V &= - DLF \cdot \Delta I^\mathrm{inj},
\end{align}
\end{subequations}
\end{minipage}
}}
\vspace{0.2cm}

\noindent
where the circuit is decomposed into a circuit $(\mathrm{pre})$ reflecting the initial network state, and a superimposed circuit $(\Delta)$ which is associated with changes in the network currents and voltages. In the considered case, these changes result solely from the DER current injections. Active and reactive power consumption of non-controllable loads is assumed to remain unchanged. Given that the initial network state is known from the state estimation result \eqref{eq:WLS_SE}, it is sufficient to solve the superimposed circuit \eqref{eq:delta_bfs} to find the new network state. Furthermore, the vector of current injections $\Delta I^\mathrm{inj}$ is \textit{sparse} due to a typically large number of nodes with no DERs. Hence, the nodes with no current injections can be removed from the model using network reduction \cite{PecenakReduction} to obtain a representation of lower dimension. Finally, the superimposed network model is used in the MPC formulation to determine the impact of the DER injection changes on the network currents and voltages at every time step $k\in\mathcal{H}$, as follows:
\scalebox{0.94}{\parbox{1.06\linewidth}{%
\begin{equation}
     \begin{bmatrix}
        \Delta I_k\\
        \Delta V_k
     \end{bmatrix} = \begin{bmatrix}
        BIBC_R \\
        -DLF_R
     \end{bmatrix}  \cdot \mathrm{diag}(1/\bar{V}_{R}^*)
            \begin{bmatrix}
                 \mathbbl{1}_{n_g} \,\,\,\, \mathbb{J}_{n_g}^* 
            \end{bmatrix}
            \begin{bmatrix}
               \Delta P_k \\
               \Delta Q_k
            \end{bmatrix}, \label{eq:BFS_final}
\end{equation}
}}
where $BIBC_R\in\{0,1\}^{N\times n_g}$ and $DLF_R\in\C^{N\times n_g}$ are reduced matrices from the BFS model \eqref{eq:bfs_branch_model}-\eqref{eq:bfs_voltage_model}, and $\bar{V}_R\in\C^{n_g}$ is the vector of known DER voltages. Active and reactive output changes of DERs are determined by subtracting the measured output values $(\bar{P},\bar{Q})\in\R^{2n_g}$ from the predicted outputs $(P_k,Q_k)\in\R^{2n_g}$ obtained in \eqref{eq:bess_p}-\eqref{eq:vshp_p}. Note that a similar result can be obtained using any other DN linear power flow method \cite{Bernstein2017} instead of the BFS.

\subsection{Bus Voltage and Thermal Loading Constraints}
Considering the dominantly resistive nature of DN lines, it is valid to assume that the angles of the bus voltages deviate only slightly from the reference angle. Therefore, it suffices to constrain the real part of bus voltages $V_{i,k}\in\C$, as follows:
\begin{equation}
	V_i^\mathrm{min} \leq \operatorname{Re}(\bar{V}_{i,k}+\Delta V_{i,k}) \leq V_i^\mathrm{max}, \quad \forall i\in\mathcal{N}, \forall k\in\mathcal{H},
\end{equation}
where $V_i^\mathrm{min}\in\R_{\geq 0}$ and $V_i^\mathrm{max}\in\R_{\geq 0}$ are the minimum and maximum allowed voltage magnitudes at every node. The thermal limit constraints are imposed by limiting the branch current magnitudes:
\begin{equation} \label{eq:current_constraint}
    |\bar{I}_{l,m,k}+\Delta I_{l,m,k}| \leq I^\mathrm{max}_{l,m} , \qquad \forall (l,m)\in\mathcal{E}, \forall k\in\mathcal{H},
\end{equation}
with $I^\mathrm{max}_{l,m}\in\R_{\geq 0}$ being the maximum admissible current for the branch connecting nodes $l$ and $m$. To preserve the linearity of the constraint set, we employ a piecewise linear approximation of the quadratic current constraints from \cite{Yang2016}. 

\subsection{Ancillary Services Delivery Constraints}
As mentioned in Sec.~\ref{sec:ASprovision}, the DN provides ancillary services by forcing active and reactive power exchanges with the transmission grid $(P_0,Q_0)$ to deviate from the scheduled values. The amount of active power to be delivered for the PFC provision is computed using the prediction model \eqref{eq:ss_general}, by simply evaluating the obtained frequency evolution against the PFC provision rule \eqref{eq:pfc}. Similarly, the amount of reactive power to be provided for voltage control depends on the voltage measurement $\bar{V}_1\in\C$ at the PCC and is computed using rule \eqref{eq:vc_q}, as follows: 
\begin{align}
    \Delta P_{k}^\mathrm{PFC} &= P_\mathrm{pfc}(\omega_k) - \bar{P}_\mathrm{DN}, \label{eq:pfc_feedback} \\
    \Delta Q_{k}^\mathrm{VC} &= Q_\mathrm{vc}(V^\star_1+ \operatorname{Re}(\Delta V_{1,k})-\|\bar{V}_1\|) - \bar{Q}_\mathrm{DN},
\end{align}
where $\bar{P}_\mathrm{DN}\in\R$ and $\bar{Q}_\mathrm{DN}\in\R$ denote the differences between the scheduled DN active and reactive power consumption and the measured consumption adjusted as a result of the ancillary services provision requirements. The voltage setpoint provided by the TSO is denoted by $V^\star_1\in\R$.
To ensure that the DN provides the required amount of PFC and VC, the power of the main feeder (i.e., the line connecting nodes $0$ and $1$) at all time steps $k$ is constrained by
\scalebox{0.88}{\parbox{1.13\linewidth}{%
\begin{equation} \label{eq:pfc_vc_provision}
    \Delta P^\mathrm{PFC}_k +j\Delta Q^\mathrm{VC}_k =V_s \cdot BIBC_{R,1} (\Delta P_{p,k} - j\Delta Q_{p,k})/\bar{V}_{R}^*,
\end{equation}
}}
with $BIBC_{R,1}\in\{0,1\}^{1\times n_g}$ denoting the row of $BIBC_R$ related to the main feeder branch, and $(\Delta P_{p,k},\Delta Q_{p,k})\in\R^{2n_g}$ being the DER output changes resulting from setpoint adjustments $(\Delta P_{p,k}^\star,\Delta Q_{p,k}^\star)$ introduced in Sec.~\ref{subsec:imcf}. The relationship between the main feeder current and power is established through the slack bus voltage $V_s\in\R$.

On the other hand, the amount of active power to be delivered for provision of SFC at each time step $k$ is obtained directly from the TSO and is denoted by $\Delta P^\mathrm{SFC}_k$. Similarly to \eqref{eq:pfc_vc_provision}, the provision of SFC is imposed by an additional change in the main feeder current, given by the following constraint:  
\begin{equation}
    \Delta P^\mathrm{SFC}_k =V_s \cdot BIBC_{R,1} \cdot \Delta P_{a,k}/\bar{V}_{R}^*, \label{eq:sfc_provision_constraint}
\end{equation}
where $\Delta P_{a,k}\in\R^{n_g}$ denotes the vector of the DER active power output changes resulting from the setpoint adjustments $\Delta P_{a,k}^\star$. It should be noted that constraints \eqref{eq:pfc_vc_provision}-\eqref{eq:sfc_provision_constraint} can be reformulated as soft constraints in order to prevent the problem from becoming infeasible in case the DN lacks resources to provide the contracted amount of reserves.

\subsection{Constrained Linear Periodic System Formulation} \label{subsec:mpcfinal}
The model presented in the previous sections is a constrained discrete-time linear time-invariant system with multirate input and a quadratic objective function. By combining \eqref{eq:bess_p}-\eqref{eq:sfc_provision_constraint} the following representation can be obtained:
\begin{align}
    x_{k+1} = A x_k + B u_k,  \qquad\forall k\in \mathcal{H},\label{eq:LTI_model}\\
    E x_k + G u_k \leq W, \qquad \forall k\in\mathcal{H}, \label{eq:LTI_constraints}
\end{align}
where $x_k\coloneqq(\Delta P_k, \Delta Q_k, \chi_k,\nu_{k})\in\R^{n_s}$ is the state vector, $n_s\coloneqq 2n_g+n_b+3n_v$ denotes the number of states, and $u_k$ is the control input defined in \eqref{eq:u_def}. The matrices $A\in\R^{n_s\times n_s}$ and $B\in\R^{n_s \times 2n_g}$ in \eqref{eq:LTI_model} are derived from the model \eqref{eq:bess_p}-\eqref{eq:vshp_p}, while $E\in\C^{n_c \times n_s},G\in\C^{n_c \times 2n_g}$ and $W\in\R^{n_c}$ are obtained from \eqref{eq:capability_curves}-\eqref{eq:sfc_provision_constraint}, with $n_c$ representing the number of constraints. Note that the above-presented model is non-minimal since the control input vector $u$ contains redundant variables - all the variables in vector $\hat{u}$ defined in \eqref{eq:u_def} that remain constant for certain time steps. The number of decision variables can thus be reduced by augmenting the state vector with the constant input vector $\hat{u}$. Let us define the periodic state $\bar{x}_k\in\R^{n_z}$  as
\begin{equation}
    \bar{x}_k\coloneqq(x_k,\hat{u}_k),\qquad n_z = \begin{cases} n_s+n_g, & \mathrm{if}\,\, z = 0 \\ n_s, & \mathrm{if}\,\, z=1
    \end{cases},
\end{equation}
where $n_z\in\mathbb{N}$ represents the number of states augmented with the number of input channels not updated at time step $k$. Furthermore, let us define $F_z\in\{0,1\}^{(m-m_z) \times n_z}$ such that $\hat{u}_k=F_z\bar{x}_k$. The system is then transformed into a linear time-varying formulation as follows:
\begin{align}
    \bar{x}_{k+1} &= \bar{A}_z \bar{x}_k + \bar{B}_z u_k \\
                  &= \bar{A}_z \bar{x}_k + \bar{B}_z(C_z\bar{u}_k+F_z\bar{x}_k) \\
                  &= (\bar{A}_z+\bar{B}_zF_z)\bar{x}_k + \bar{B}_zC_z\bar{u}_k, \label{eq:sys_final}
\end{align}
with $\bar{A}_z\in\R^{n_z \times n_z}$ and $\bar{B}_z\in\R^{n_z\times 2n_g}$ denoting the periodic state and control matrices, defined by:
\begin{align}
    \bar{A}_z &= A,\quad\qquad\quad\,\,\, \bar{B}_z = B,\qquad\qquad\,\, \mathrm{if}\,\, z=1, \\
    \bar{A}_z &= \begin{bmatrix}
        A & \mathbbl{0}_{n_g} \\ 
        \mathbbl{0}_{n_g} & \mathbbl{1}_{n_g}
    \end{bmatrix},
    \bar{B}_z = \begin{bmatrix}
        B & \mathbbl{0}_{n_g} \\ 
        \mathbbl{0}_{n_g} & \mathbbl{1}_{n_g}
    \end{bmatrix}, 
    \mathrm{if}\,\, z = 0.
\end{align}
Note that the matrix $C_z$ was previously introduced in \eqref{eq:Cz}. 

The inequality constraints can be transformed by applying a similar procedure, as follows:
\begin{align}
    \bar{E}_z\bar{x}_k + Gu_k \leq W, \\
    (\bar{E}_z+GF_z)\bar{x}_k + GC_z\bar{u}_k\leq W, \label{eq:constr_final}
\end{align}
where $\bar{E}_z\in\R^{n_c\times n_z}$ takes into account the periodic property of the state vector. The final model described by \eqref{eq:sys_final} and \eqref{eq:constr_final} defines a constrained linear periodic system. Together with the objective function \eqref{eq:objective_fcn} a quadratic optimization problem in variables $\{\bar{u}_k,\forall k\in\mathcal{H}\}$ is obtained, which can be solved efficiently using modern optimization solvers. 

\section{Results} \label{sec:res}
\begin{figure}[!t]
    \centering
    \includegraphics[scale=0.525]{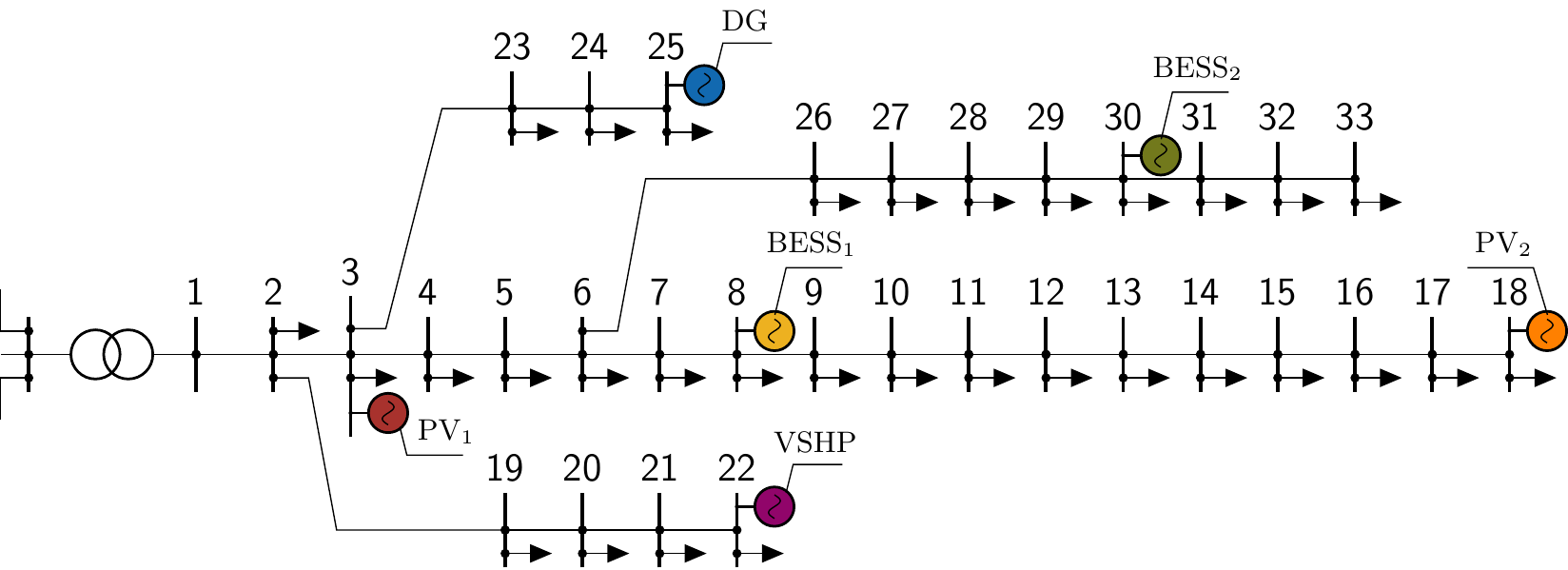}
    \caption{Customized IEEE 33-bus system, with the DER units placed at the following nodes: $3,8,18,22,25$ and $30$.}
    \label{fig:33busIEEE}
\end{figure}

In this section, the proposed DN controller is implemented and examined on a modified version of the IEEE 33-bus network \cite{Baran1989}, shown in Fig.~\ref{fig:33busIEEE}. The system has been customized by adding PV units at nodes $\mathcal{P}=\{3,18\}$, BESS at nodes $\mathcal{B}=\{8,30\}$, a DG at node $\mathcal{D}=\{25\}$, and a VSHP at node $\mathcal{V}=\{22\}$. The DG rated power is set to \SI{670}{\kilo\voltampere}, with the diesel governor and the excitation system time constants of $T_g = \SI{10}{\second}$ and $T_e = \SI{1}{\second}$, respectively. It is set to operate at its minimum allowable power output of \SI{100}{\kilo\watt}. The two BESS are identically parametrized with \SI{500}{\kilo\voltampere} rated power and storage capacity of BESS \SI{160}{\kWh}. Initial active and reactive power setpoints are set to zero. The two PV units operate at $90\%$ of their respective peak powers of \SI{150}{\kilo\watt} and \SI{300}{\kilo\watt}. The VSHP operates with a unity power factor and at the active power setpoint point of $\SI{200}{\kilo\watt}$. The total load consumption of the network amounts to \SI{3.9}{\mega\watt}, which is assumed to lead to nominal thermal loading of the grid. The minimum and maximum acceptable voltages at each bus are set to $0.9$ and $1.1$ p.u., respectively, and the current limit of each line is set to $120\%$ of the nominal thermal loading. The per-unit system used throughout this section is defined by the base power of \SI{1}{\mega\watt} and base voltage of \SI{12.66}{\kilo\volt}.

All simulations are performed using a comprehensive distribution network DAE modelling framework developed in-house. The model includes detailed representation of relevant dynamics and controls pertaining to individual DERs as well as the dynamics of distribution lines and loads. The network lines are modeled as $\pi$-sections \cite{UrosStability}, the low voltage feeders as composite loads, and the transmission network as a Th\'evenin equivalent with a controllable frequency voltage source. For the diesel generator, we consider a $5^\mathrm{th}$ order synchronous machine with $2^\mathrm{nd}$ order diesel governor and an excitation system including a reactive power control loop. Furthermore, PV and BESS unit AC-sides are represented by \textit{grid-feeding} inverters \cite{Rocabert2012}, including a phase-locked loop, a power measurement, a current control loop, and an averaged switching unit. The dynamic model of the BESS presented in \cite{BESS2019} was adopted. The considered VSHP model includes a rectifier and an inverter with their respective controls \cite{Johanna2020}. Furthermore, the inverter is connected to an induction machine that drives the shaft of the heat pump’s compressor \cite{IbrahimHeatPumps2020}.

It is assumed that the DN has sold $\bar{Q}_\mathrm{vc}=0.5\,\mathrm{p.u.}$ of the voltage control reserve, $\bar{P}_\mathrm{pfc} = 1\,\mathrm{p.u.}$ of the primary control reserve, and $1\,\mathrm{p.u.}$ of the secondary control reserve in the ancillary service market, which are to be supplied according to the dispatch rules introduced in Sec.~\ref{sec:ASprovision}. The controller operates at a rate of $T_s = T_{s,p} = \SI{1}{\second}$ to account for the time needed to compute the optimal setpoints and communicate them to the individual DERs. On the other hand, the SFC signal is assumed to be received at a significantly slower rate of $T_{s,a} = \SI{10}{\second}$. The control horizon consists of $30$ time steps (or horizon length of \SI{30}{\second}), which corresponds to three time steps for the SFC-related variables. Potential issues associated with the communication infrastructure, such as delays and failures, are not considered since the focus of this work is on the control design.  Modelling of the optimization procedure was performed using YALMIP \cite{Lofberg2004}, while GUROBI was used as the solver. The average solver time required for the solution of the algorithm is around \SI{100}{\milli\second}, with the computations performed on an Intel i9-8850H processor.

In the rest of this section, we first evaluate the modelling error introduced by the power flow linearization and network decomposition. Subsequently, we present the simulation results for two case studies including frequency and voltage disturbances in the transmission system leading to deviations that necessitate the activation of the ancillary services. 

\subsection{Superimposed Circuit Linearization Error Analysis}
\begin{figure}[!b]
    \centering
    \includegraphics[scale=0.61]{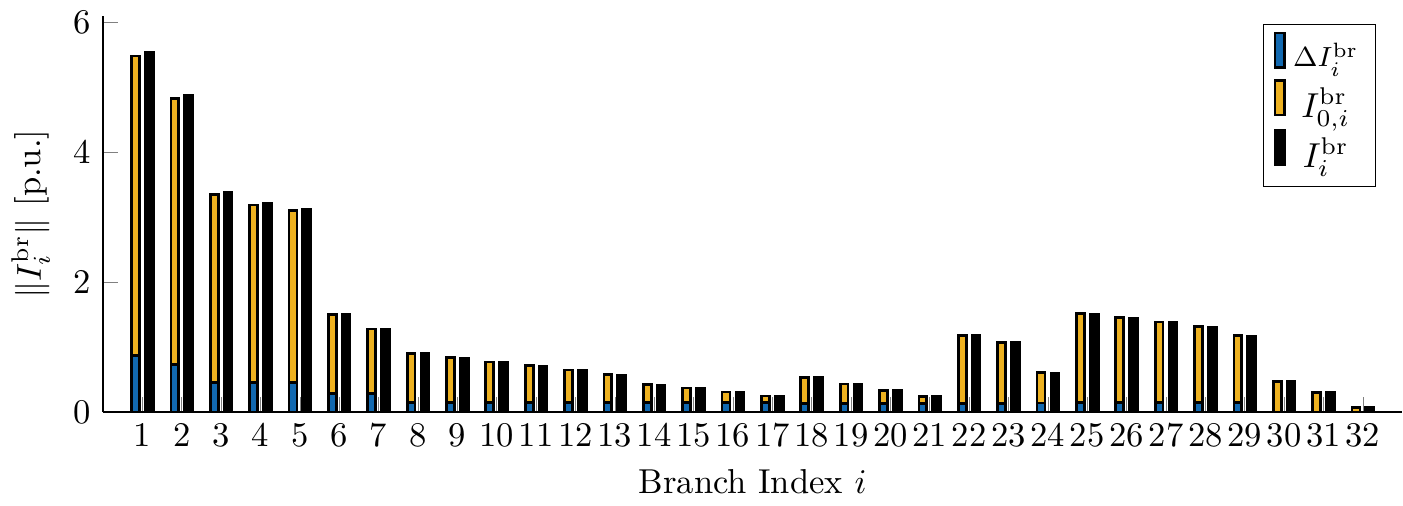}
    \caption{An example of branch current computation using the superimposed circuit principle. Individual branches are indexed by their receiving end nodes.}
    \label{fig:superimposed_branch_res}
\end{figure}

To illustrate the superimposed circuit principle introduced in Sec.~\ref{subsec:BFSdecomposed} we perform a case study on the previously presented modified version of the IEEE 33-bus system. An initial grid operating point is assumed to be known and corresponds to the nominal load consumption (as given in \cite{Baran1989}) and zero DER output. A change in the DER power output of $\Delta P_d = \SI{0.1}{\mathrm{p.u.}}$, $\Delta Q_d = \SI{0.1}{\mathrm{p.u.}}, \forall d\in\mathcal{R}$ is imposed on the system. Simulation results shown in Fig.~\ref{fig:superimposed_branch_res} illustrate the results of the computation of branch currents using the superposition principle. The initial operating point and the superimposed circuit solution \eqref{eq:delta_bfs}, denoted respectively by $I_0^\mathrm{br}$ and $\Delta I^\mathrm{br}$, together approximate the nonlinear (Newton-Raphson-based) power flow solution indicated by $I^\mathrm{br}$. As can be seen from the figure, the superimposed circuit model provides a good quality approximation of the nonlinear power flow solution since only a minor error is introduced.
\begin{figure}[!t]
    \centering
    \includegraphics[scale=0.8]{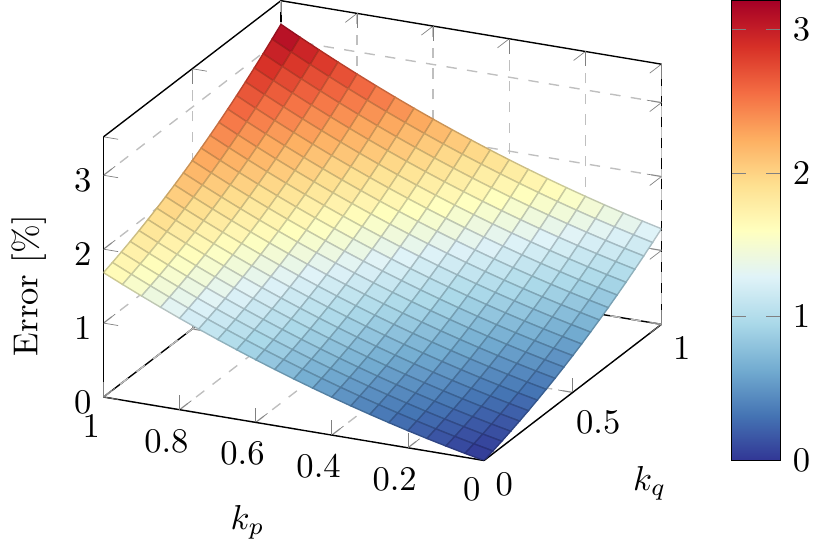}
    \caption{Relative linearization error of voltage magnitudes as a function of DER active and reactive power injections.}
    \label{fig:superimposed_err_res}
\end{figure}

Furthermore, the error introduced by the approximation can be quantified by continuation analysis. Let $\Delta P_d = k_p P_d^\mathrm{max}$ and $\Delta Q_d = k_q Q_d^\mathrm{max}$ for all $d\in\mathcal{R}$, where $P_d^\mathrm{max}$ and $Q_d^\mathrm{max}$ are the maximum allowable power outputs of the corresponding DER units. Parameters $k_p$ and $k_q$ are independently swept through the interval $[0,1]$ with granularity of $0.01$. The metrics of the relative error defined by $\|(|V_0+\Delta V| - |V|)/|V|\|_2$ are used to quantify the approximation error, where $V_0$ indicates voltages corresponding to the initial operating point, $\Delta V$ is the solution to \eqref{eq:delta_bfs}, and $V$ denotes the Newton-Raphson solution. The error surface depicts the results of the analysis in Fig.~\ref{fig:superimposed_err_res}, which indicates an error of around $3\%$ for the unlikely scenario where all DERs are ramped up from 0 to their maximum power outputs. Small setpoint changes in the blue region, where the linearization error is below $1\%$, can be expected during normal controller operation.

\subsection{A Loss of Generation Event}
\begin{figure}[!b]
    \centering
    \includegraphics[scale=1.175]{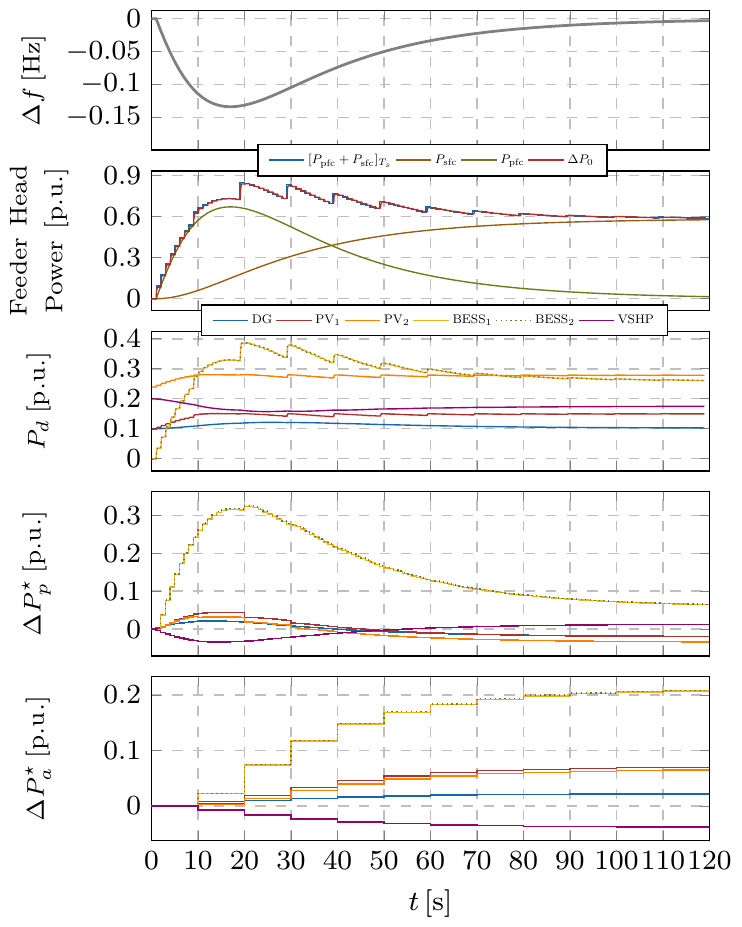}
    \caption{Time-domain response after a generation loss: (i) frequency deviation; (ii) DN power output deviation and frequency control references; (iii) DER active power outputs; (iv) PFC setpoint adjustments; and (v) SFC setpoint adjustments.}
    \label{fig:gen_loss_pf}
\end{figure}
In this section, the controller performance under a power imbalance event in the form of a generator loss is analyzed. The event leads to a frequency decline as shown in Fig.~\ref{fig:gen_loss_pf}, which necessitates activation of PFC and SFC reserves and the appropriate controller action. The reserve provision plot, i.e. the second plot in Fig.~\ref{fig:gen_loss_pf}, showcases the DN active power schedule deviation $\Delta P_0$, and PFC and SFC provision requirements computed using \eqref{eq:pfc} and \eqref{eq:sfc}. As can be observed, the DN under the proposed control scheme timely and accurately supplies the required active power according to the ancillary service provision requirements. 
Furthermore, outputs of individual units are presented in the third plot. As can be seen, the main reserve provision carriers are BESS, which are the most flexible units capable of fast ramping. The two BESS are dispatched identically, suggesting that their different placement in the network has no influence on the reserve provision in this case study. On the other hand, the contribution of the two PV units is minor due to the limited upward flexibility. Furthermore, output of $\mathrm{PV}_2$ is constrained by a branch capacity limit as will be shown later.  
The VSHP participates by reducing its power output up to $60\%$ of its current consumption, with slower ramping compared to PV and BESS units. Lastly, the large governor constant together with the high fuel costs of the DG unit limits its participation in the services provision. 

Due to the modular controller structure, the setpoint changes for PFC and SFC provision are computed separately and also presented in Fig.~\ref{fig:gen_loss_pf}. Nonetheless, the total setpoint change applied to DERs at each time step is the sum of individual setpoint changes, as indicated in \eqref{eq:u_def}. Interdependencies between the provision of the individual services exist since the same network resources are used for the provision of all services. Namely, the PFC setpoints of the two PV units are reduced to a negative value at $t>\SI{40}{\second}$ to open up capacity for the SFC provision. The redistribution of the setpoints is governed by the selection of the costs in objective function \eqref{eq:objective_fcn}. Finally, it is also worth noting that since VSHPs are loads consuming active power, their negative setpoint change results in consumption reduction and correspondingly a positive contribution towards PFC and SFC provision.

\begin{figure}[!b]
	\centering
	\begin{minipage}{0.45\textwidth}
		\centering
		\hspace{-0.4cm}
		\scalebox{1.125}{\includegraphics[]{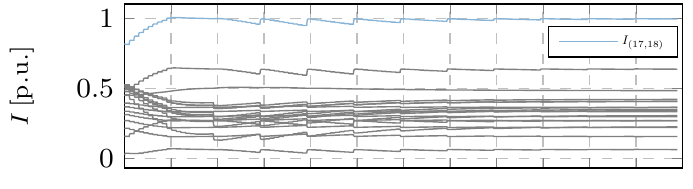}}\\
		\vspace{-0.1cm}  
	\end{minipage}
	\begin{minipage}{0.45\textwidth}    
		\centering
		\hspace{-0.15cm}
		\scalebox{1.125}{\includegraphics[]{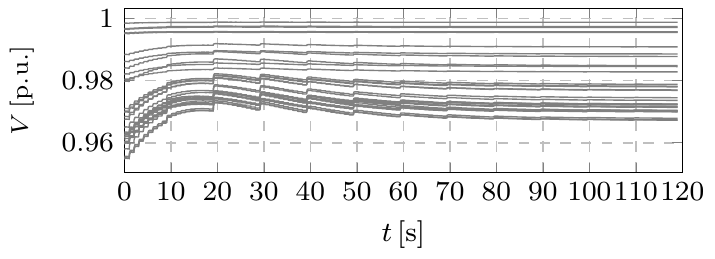}}\\
	\end{minipage}
	\caption{\label{fig:gen_loss_network} Evolution of branch currents (top) and bus voltages (bottom) during the controller operation. Branch currents are normalized by their thermal limits.}
\end{figure}
The behavior of the DN bus voltages and branch currents is presented in Fig.~\ref{fig:gen_loss_network}. The DN is deloaded due to the voltage response of the loads and combined with the increased active power injection due to the PFC and SFC provision, leads to increasing voltage values. On the other hand, branches that were subject to reverse power flow due to high PV injections are becoming overloaded, with branch $(17,18)$ reaching its capacity limit. Hence, the PV at node 18 is not able to provide more power due to the network constraint.

\begin{figure}[!t]
    \centering
    \includegraphics[scale=1.125]{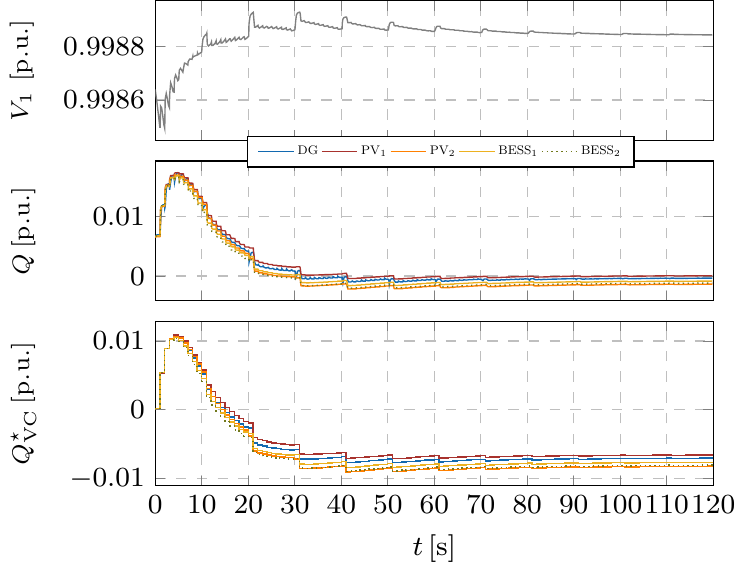}
    \caption{Time-domain response of voltage control-related variables: (i) voltage magnitude at bus 1; (ii) DER reactive power outputs; (iii) voltage control setpoint adjustments.}
    \label{fig:gen_loss_vc}
\end{figure}

Lastly, the voltage control-related quantities are presented in Fig.~\ref{fig:gen_loss_vc}. On the transmission level, the loss of a generator is accompanied by depressed voltages. Therefore, the bus $1$ voltage starts to drop after the fault and appropriate action of the controller follows, injecting reactive power to prevent the voltage decline. However, the deloading of the DN resulting from PFC and SFC-related DER active power injections soon becomes dominant and raises the voltage magnitude. The reactive power consumption now needs to be increased to prevent the excessive voltage rise. To this end, negative setpoint changes are being applied to all DERs. Finally, the controlled voltage magnitude at bus $1$ settles at a steady state governed by \eqref{eq:vc_q}.  
\begin{figure}[!t]
	\centering
	\begin{minipage}{0.45\textwidth}
		\centering
		\hspace{-0.15cm}
		\scalebox{1.12}{\includegraphics[]{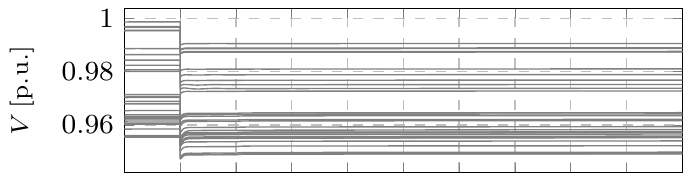}}\\
		\vspace{-0.3cm}  
	\end{minipage}
	\begin{minipage}{0.45\textwidth}    
		\centering
		\hspace{-0.4cm}
		\scalebox{1.12}{\includegraphics[]{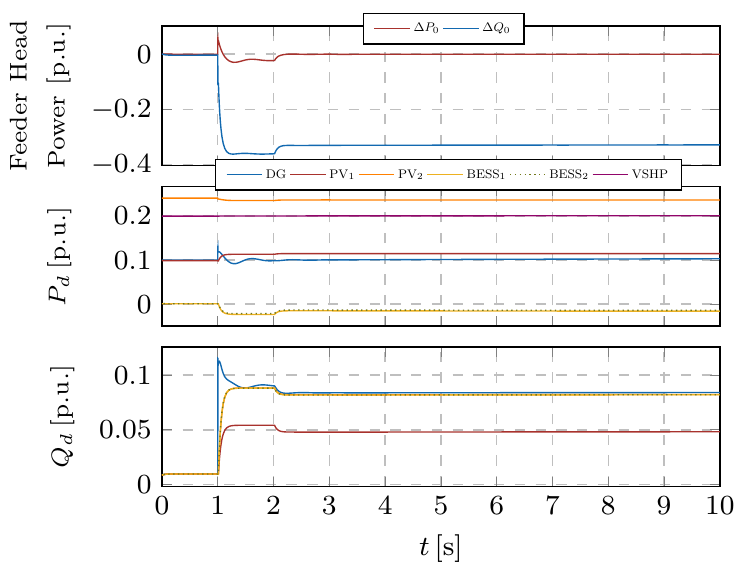}}\\
	\end{minipage}
	\caption{\label{fig:line_trip_general}Time-domain response of the network after a line trip event: (i) bus voltage magnitudes; (ii) main feeder active and reactive power deviations; (iii) DER active power output; (iv) DER reactive power output.}
\end{figure}

\subsection{Line Trip Event}
To assess the controller performance under voltage disturbances, we simulate a line trip event by modifying the Th\'evenin impedance of the transmission network equivalent. A step-wise change of $0.1\,\,\mathrm{p.u.}$ is applied to the resistive part of the impedance and a consequent drop in voltage magnitudes at all busses occurs, as shown in Fig.~\ref{fig:line_trip_general}. The second plot showcases active and reactive power deviations from the scheduled values at the feeder head. The controller responds promptly by injecting around $0.4\,\,\mathrm{p.u.}$ of reactive power in response to the voltage drop. A minor overshoot occurs due to modeling errors. Nevertheless, it is rapidly corrected in the next time step. Individual units react with similar control effort due to equal cost coefficients for reactive power provision in \eqref{eq:objective_fcn}. Furthermore, since the voltage drop event has modified the active power exchange at the PCC, the controller adjusts the DER active power setpoints to bring the feeder power flow back to the scheduled value. This functionality is a result of \eqref{eq:pfc_feedback} and \eqref{eq:pfc_vc_provision}, which ensure compensation of any power deviation at the PCC that is not the result of the frequency control provision. 

\section{Conclusion} \label{sec:concl}
This paper proposes a novel centralized controller to aggregate and dispatch DERs in an active DN for the provision of voltage support and primary and secondary frequency control. At the heart of the controller lies a multi-rate MPC scheme capable of accommodating distinct timescales and provision requirements of each ancillary service and ensuring that the available resources are properly allocated. Additionally, a network decomposition approach applied to a linear power flow model is considered to reduce the problem dimensionality and thus improve the computational efficiency of the controller. Numerical simulations revealed that the proposed decomposition method introduces only a low error (below $3\%$) while significantly improving the computation times. Furthermore, two case studies were presented to demonstrate the controller performance. It is found that the controller successfully allocates power setpoint changes to DERs in real-time and accurately provides multiple ancillary services to the transmission system. The limitations of the DN as an ancillary service provider are reflected in the preexisting bottlenecks in the grid. Moreover, the placement of DERs in the network also might limit resource utilization. The main drawback of the controller is its centralized implementation which is susceptible to single point failure and communication issues. In future work, we will consider the viability of distributed implementations of the proposed controller.

\bibliographystyle{IEEEtran}
\bibliography{bibliography}

\end{document}